\documentclass{webofc}

\RequirePackage{lineno}
\usepackage[varg]{txfonts}   
\usepackage[utf8]{inputenc}
\usepackage{ulem}
\usepackage{amssymb}
\usepackage{xspace} 
\usepackage{amsmath}
\usepackage{multirow}
\usepackage{float}
\usepackage{harpoon}
\usepackage{MnSymbol}
\usepackage{appendix}
\usepackage{color}
\RequirePackage{lineno} 



\makeatletter 
\@addtoreset{equation}{section}
\makeatother  

\newcommand{\Yphi}{$Y(\Delta\phi)$}

\newcommand{\sqrtsNN}{\mbox{$\sqrt{s_{\mathrm{NN}}}$}}

\newcommand{\pT} {p_{\mathrm{T}}}

\newcommand{\Deta}{\mbox{$\Delta \eta$}}

\newcommand{\heau}{\mbox{$^{3}$He$+$Au}\xspace}
\newcommand{\dau}{\mbox{$d$$+$Au}\xspace}

\newcommand{\pau}{\mbox{$p$$+$Au}\xspace}
\begin{document}
\title{Measurements of azimuthal anisotropies in $^{16}$O+$^{16}$O and $\gamma$+Au collisions from STAR}
%
%

\author{\firstname{Shengli} \lastname{Huang}\inst{1} \fnsep\thanks{\email{shengli.huang@stonybrook.edu}} for the STAR Collaboration 
}

\institute{ Stony Brook University, Chemistry Department}

\abstract{%

In these proceeding, we present the first measurements of azimuthal anisotropies, $v_2$ and $v_3$, in $^{16}$O+$^{16}$O collisions at 200 GeV as a function of transverse momentum and multiplicity, by using two- and four-particle correlation methods. We compare our measurements with STAR measurements of $v_n$ in \dau and \heau collisions to provide insight into the impact of system symmetry on initial condition for small systems. We also investigate the ratio $v_2\{4\}/v_2\{2\}$ as a function of centrality, which is expected to be sensitive to nucleon-nucleon correlation in the $^{16}$O nucleus. 
}
\maketitle
\section{Introduction}
\label{intro}
Recently, the anisotropic flow harmonics have been extensively measured in various small system collisions via two- and multi-particle correlations from $p$+$p$~\cite{CMS:2010ifv,ATLAS:2015hzw} to $p$+A~\cite{CMS:2012qk,ALICE:2012eyl,Aad:2012gla,PHENIX:2022nht,STAR:2022pfn}, and $\gamma$+A collisions~\cite{ATLAS:2021jhn}.
However, the origin of collectivity in small system collisions still lacks satisfactory explanations, primarily due to the relatively limited understanding of the initial conditions in small systems. The initial geometry in small systems is predominantly influenced by fluctuations, encompassing not only position fluctuations from nucleons and sub-nucleons but also longitudinal dynamical fluctuations~\cite{Huang:2019tgz}. Moreover, nucleon-nucleon correlations, such as nucleonic clusters in light nuclei, can also significantly impact the initial geometry~\cite{PhysRevLett.112.112501,Ma:2022dbh}. 
The small system collision scan at RHIC, including both symmetric and asymmetric small systems (O+O $>$ \heau $>$ \dau $>$ \pau $>$ $\gamma+$Au), could provide a better understanding of initial conditions.

\section{Measurements of di-hadron correlations in $^{16}$O+$^{16}$O collisions}
\label{sec-1}

The charged hadrons are detected in the Time Project Chamber (TPC) ~\cite{Wang:2017mbk} 
at STAR detector which covers the pseudo-rapidity range around $|\eta|\leq 1.5$. 
The per-trigger yield of two-particle azimuthal angular correlations \Yphi\ = $1/N_{\mathrm{Trig}}dN/d\Delta\phi$ is measured to extract the anisotropy harmonics. The two-track efficiency corrections are evaluated via single-particle efficiency from embedding in peripheral Au + Au collisions. 
%

\begin{figure*}[ht]
\begin{center}
\sidecaption
  \includegraphics[width=0.6\linewidth]{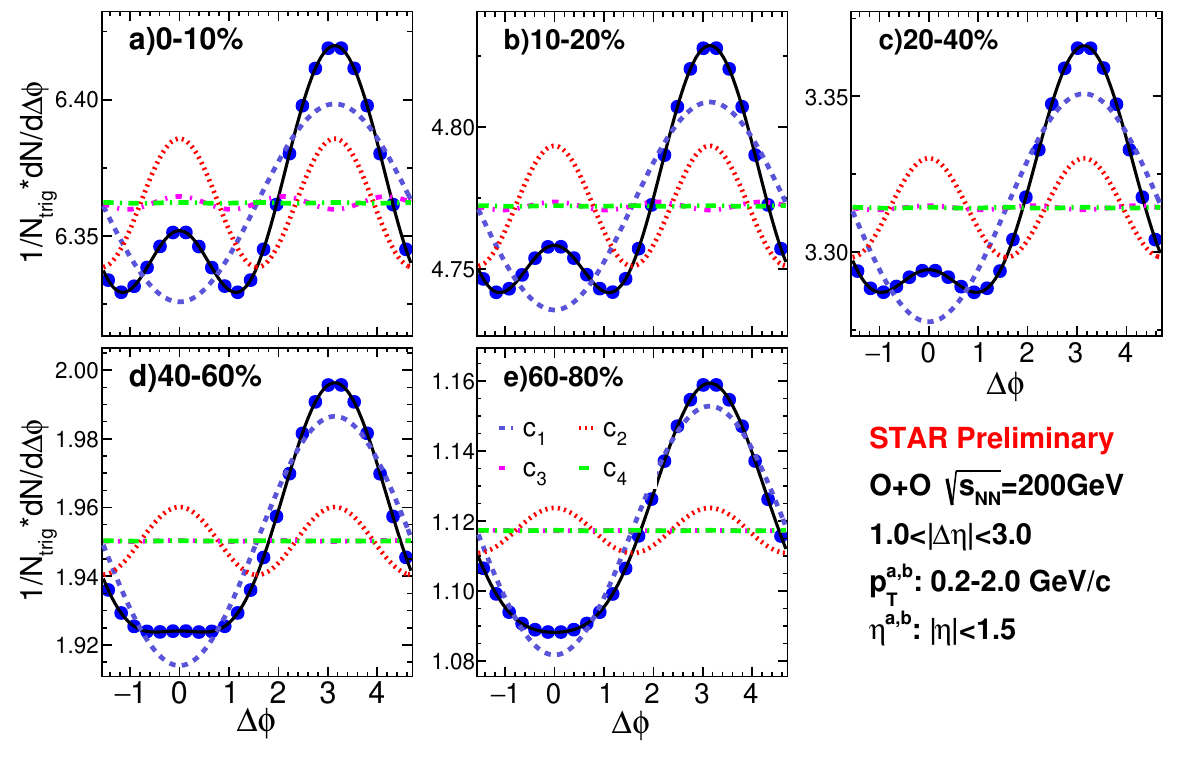}
  \caption{
  Two-particle per-trigger yield distributions in $^{16}$O+$^{16}$O collisions collisions at $\sqrtsNN$ = 200 GeV for different centralities; the trigger and associated particles are selected in $0.2 < \pT < 2.0$~GeV/c and 1.0 $< |\Deta| <$ 3.0. An illustration of the Fourier functions fitting procedure, to estimate the ``nonflow" contributions and extract the $v_2$ and $v_3$ flow coefficients, is also shown.
}
 \label{fig:corr_functions}
 \end{center}
\end{figure*}

Figure~\ref{fig:corr_functions} shows the distributions \Yphi\ for $^{16}$O+$^{16}$O collisions in different centralities. The centrality here is defined with total multiplicity measured with Event-Plane-Detector (EPD) 
, which covers $2.1<|\eta|<5.1$. For these correlators, the trigger ($\mathrm{Trig}$)- and the associated (Assoc)-particles are measured in the range $0.2 < \pT < 2.0$~GeV/c and 1.0 $< |\Deta| <$ 3.0. The near- and away-side patterns of the distributions for central $^{16}$O+$^{16}$O collisions indicate a sizable influence from flow, and ``nonflow" correlations that can be removed with the subtraction methods outlined below. The correlator for 60-80\%  $^{16}$O+$^{16}$O collisions (Fig.~\ref{fig:corr_functions}(e)) is dominated by ``nonflow" correlations, and thus can be used to estimate ``nonflow" contributions in central $^{16}$O+$^{16}$O collisions. 

A Fourier function fit is employed to the measured \Yphi\ distributions to extract $v_{2,3}(\pT^{\mathrm{Trig.}})$ as:

\begin{equation}
 Y(\Delta\phi, \pT^{\mathrm{Trig.}}) = c_{0}(1 + \sum_{n=1}^{4} \, 2c_{n}\, \cos ( n \,\Delta\phi )). 
 \label{eq:fourier}
 \end{equation}

where $c_{0}$ represents the average pair yield (also referred to as the pedestal), and $c_n$ (for $n=1$ to 4) are the Fourier coefficients. The corresponding harmonic components are depicted by the colored dashed lines in Fig.~\ref{fig:corr_functions}.
The non-flow contributions are subtracted with:
\begin{equation}
c_{n}^{\mathrm{sub}} = c_{n} - c^{nonflow}_{n} = c_{n} - c_{n}^{peri.}\times f
\label{eq:c2_collective_noncollective}
 \end{equation}
where the $c_{n}^{\mathrm{sub}}$ is $c_{n}$ after nonflow subtraction. The methods differ from each other in terms of how the scale factor $f$ is estimated.
Four established methods are implemented to estimate the factor $f$ with the details which can be found in ref.~\cite{STAR:2022pfn}. Systematic uncertainties account for the variations among the four 
methods.


The $c_n$ is simply the product of $v_n$ for trigger- and associated-particles, i.e. $c_{n}=v_{n}^{\mathrm{Trig.}} \times v_{n}^{\mathrm{Assoc.}}$

\section{$v_{n}$ in symmetric and asymmetric small systems}
\begin{figure*}[ht]
\centering
\sidecaption
\includegraphics[width=0.6\linewidth]{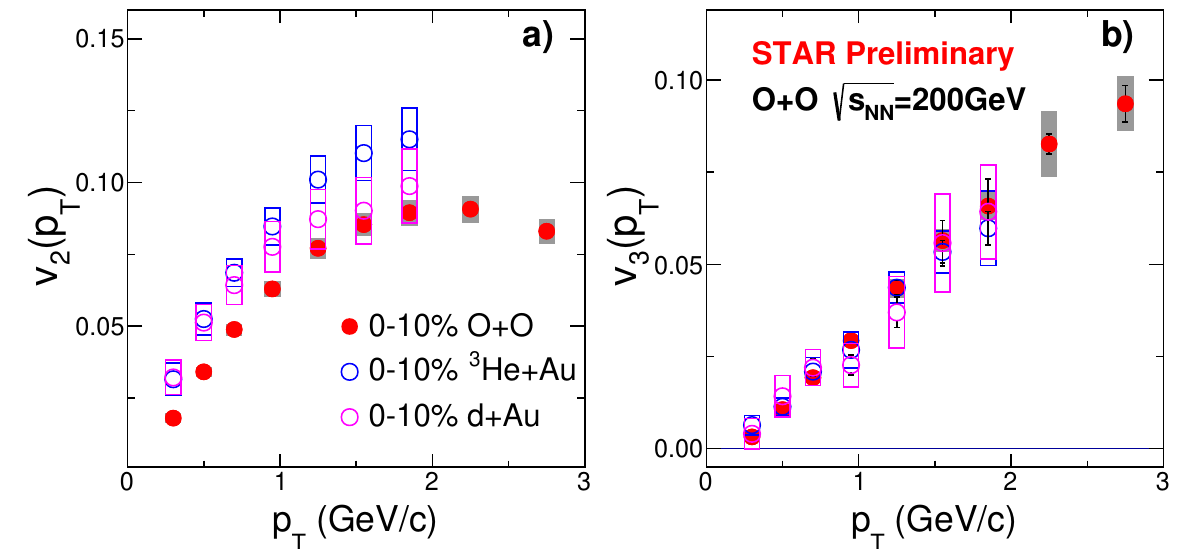}
\caption{The $v_2(\pT)$ values (left panels) and $v_3(\pT)$ values (right panels) in the 0-10\% $^{16}$O+$^{16}$O and compared with that in 0-10\% \dau and \heau collisions}
\label{fig-2}       
\end{figure*}

The $v_{2}(p_{T})$ and $v_{3}(p_{T})$ in 0-10\% $^{16}$O+$^{16}$O collisions are compared with that in 0-10\% \dau and \heau collisions as shown in the Figure.~\ref{fig-2}. As shown in panel (a), the $v_{2}(p_{T})$ in 0-10\% $^{16}$O+$^{16}$O is smaller than that from \dau and \heau collisions. However, the values of $v_{3}(p_{T})$ shown in panel (b) are similar among the three small systems. It is consistent with the initial geometry predicted by Glauber 
model calculations, which include sub-nucleon fluctuations~\cite{Welsh:2016siu}. In such a model, the $\varepsilon_{2}$ are similar between \dau and \heau collision and larger than that of $^{16}$O+$^{16}$O collisions, while $\varepsilon_{3}$ are similar between three systems.

\section{Centrality dependence of $v_{2}\{4\}/v_{2}\{2\}$ in $^{16}$O+$^{16}$O collisions }

Protons and neutrons can organize themselves into sub-group structures known as clusters within nuclei. In nuclei such as $^{16}$O with double magic numbers—where the neutron and proton (atomic) numbers each equals—two protons and two neutrons exhibit a tendency to group together, forming a alpha cluster~\cite{Furutachi:2007vz}.

The impact of clusters on the initial geometry fluctuations differs significantly from the predictions of two major \textit{ab initio} ~\cite{Meissner:2014lgi} methods. One approach stems from nuclear lattice effective field theory (NLEFT)~\cite{Elhatisari:2017eno}, while the other involves quantum Monte Carlo calculations utilizing chiral effective field theory Hamiltonians (VMC)~\cite{Gezerlis:2013ipa}. Consequently, measuring the initial geometry fluctuations in $^{16}$O+$^{16}$O collisions becomes essential for gaining insights into nucleon-nucleon correlation and for constraining the varied predictions of the \textit{ab initio} lattice effective field theory.

The initial geometry fluctuation can be measured via the ratio of $v_{2}\{4\}/v_{2}\{2\}$~\cite{Bilandzic:2010jr}, where
\begin{equation}
\begin{aligned}
    v_{2}\{2\}^{2} &= \left\langle v_{2}^{2}\right\rangle\\
    v_{2}\{4\}^{4} &= 2\left\langle v_{2}^{2}\right\rangle^{2}-\left\langle v_{2}^{4}\right\rangle
\end{aligned}
\end{equation}
\noindent since the initial geometry has a strong linear relation with final state, i.e. $\varepsilon_{2}\{4\}/\varepsilon_{2}\{2\}$= $K \times v_{2}\{4\}/v_{4}\{2\}$,  where $K$ captures the response from medium dynamical properties.
  
\begin{figure*}
\centering
\sidecaption
\includegraphics[width=0.6\linewidth]{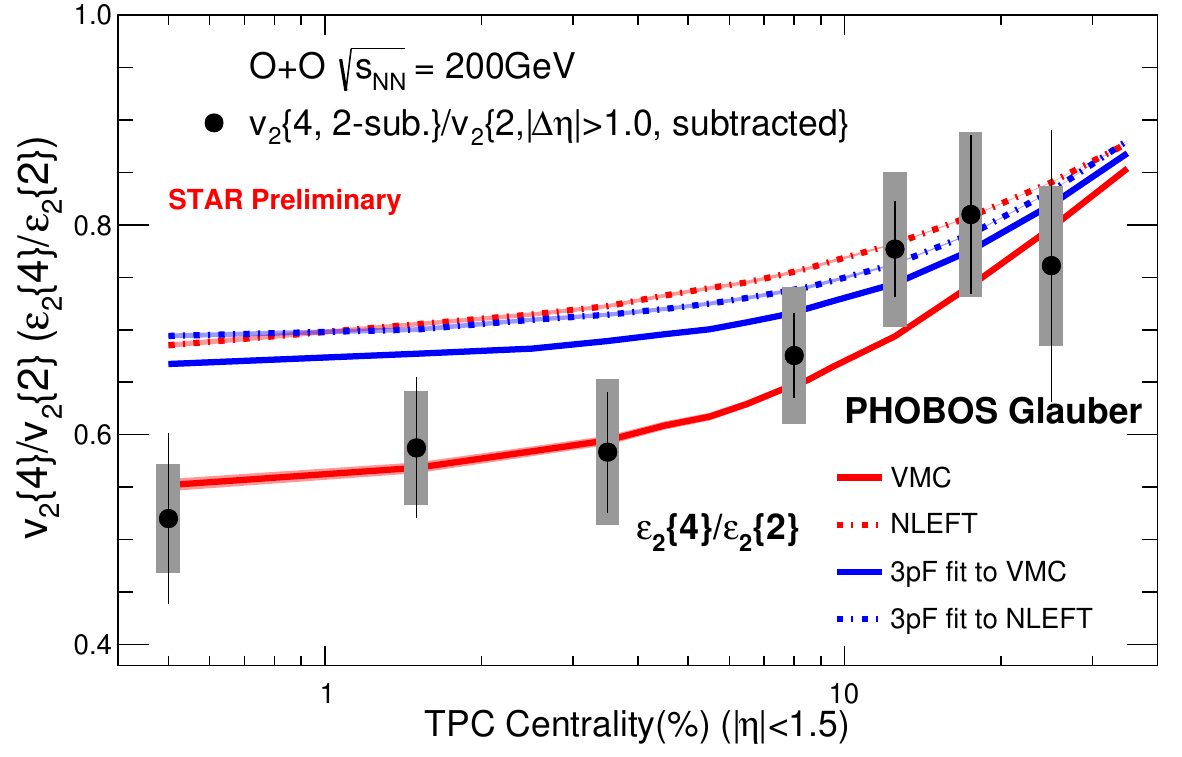}
\caption{
The figure illustrates $v_{2}\{4\}/v_{2}\{2\}$ as a function of centrality, defined by charged hadron multiplicity at $|\eta|<1.5$, in $^{16}$O+$^{16}$O collisions. Additionally, the $\varepsilon_{2}\{4\}/\varepsilon_{2}\{2\}$ ratio from NLEFT, VMC, and two types of 3pF distributions are presented for comparison.
Note that an issue is identified in the publicly available PHOBOS Glauber, which affected the implementation of the NLEFT and VMC configuration. This has been corrected in the updated figure}
\label{fig-3}       
\end{figure*}

Figure~\ref{fig-3} depicts the ratio $v_{2}\{4\}/v_{2}\{2\}$ as a function of centrality, defined by charged hadron multiplicity measured at $|\eta|<1.5$. The $\varepsilon_{2}\{4\}/\varepsilon_{2}\{2\}$, calculated using the PHOBOS Glauber model~\cite{PHOBOSGlauber} with $^{16}$O configurations from NLEFT, VMC models and three-parameter Fermi (3pF) distribution which fits to the radial density distribution from aforementioned models respectively, is also presented for comparison. It is noteworthy that we identified an issue in the public PHOBOS Glauber code related to the implementation of $^{16}$O configurations, and we have since rectified it. Consequently, the calculation presented here differs from that showcased in the QM presentation.

Upon comparison, our findings indicate that the measurements align more closely with the eccentricity ratio from the VMC model, whereas they are considerably smaller than those from the NLEFT model or 3pF distributions. Nevertheless, a detailed hydrodynamics model and transport model are imperative to determine the parameter $K$. It will further decipher the difference and help to constrain the test of the performance between different \textit{ab initio} models.

\section{Summary}
We compare the measured $v_2(\pT)$ and $v_3(\pT)$ in 0-10\% $^{16}$O+$^{16}$O collisions at $\sqrt{s_{NN}} =$ 200 GeV with those in 0-10\% \dau and \heau collisions. This comparison underscores the significance of sub-nucleon fluctuations in small systems. The ratio $v_{2}\{4\}/v_{2}\{2\}$ is observed to be closer to the $\varepsilon_{2}\{4\}/\varepsilon_{2}\{2\}$ ratio from the VMC calculation, while being smaller than that from NLEFT. This observation suggests that $v_{2}\{4\}/v_{2}\{2\}$ can serve as a powerful tool for studying nucleon-nucleon correlations in collisions involving light nuclei.

Looking ahead, the measurements of $\gamma$+Au collisions from the Au+Au data taken in 2021 and 2023 will provide further insights into understanding initial conditions such as sub-nucleon fluctuations and nucleon-nucleon correlations.

This work is supported by the U.S. Department of Energy, Award number DE-SC0024602.



%
\bibliography{ref}
%
%
%
%

\end{document}